\documentclass[superscriptaddress,
reprint,
 amsmath,amssymb,
 aps,
]{revtex4-2}

\usepackage{graphicx}
\usepackage{dcolumn}
\usepackage{bm}
\usepackage{hhline}
\usepackage{braket}
\usepackage[separate-uncertainty=true]{siunitx}
\DeclareSIUnit{\sample}{S}
\DeclareSIUnit{\dBm}{dBm}
\usepackage{xcolor}

\setcitestyle{super}

\DeclareMathOperator{\atantwo}{arctan2}

\usepackage[normalem]{ulem}

\begin{document}

\renewcommand{\s}[1]{$\lvert\textrm{S{#1}}\rvert$}

\title{Coherent control of magnon-polaritons using an exceptional point}

\author{N. J. Lambert}
\email{nicholas.lambert@otago.ac.nz}
\affiliation{Department of Physics, University of Otago, Dunedin, New Zealand}
\affiliation{The Dodd-Walls Centre for Photonic and Quantum Technologies, New Zealand}
\author{A. Schumer}
\affiliation{Institute for Theoretical Physics, Vienna University of Technology (TU Wien), A-1040 Vienna, Austria}
\author{J. J. Longdell}
\affiliation{Department of Physics, University of Otago, Dunedin, New Zealand}
\affiliation{The Dodd-Walls Centre for Photonic and Quantum Technologies, New Zealand}
\author{S. Rotter}
\affiliation{Institute for Theoretical Physics, Vienna University of Technology (TU Wien), A-1040 Vienna, Austria}
\author{H. G. L. Schwefel}
\affiliation{Department of Physics, University of Otago, Dunedin, New Zealand}
\affiliation{The Dodd-Walls Centre for Photonic and Quantum Technologies, New Zealand}

\date{\today}
\begin{abstract}
    The amplitude of resonant oscillations in a non-Hermitian environment can either decay or grow in time, corresponding to a mode with either loss or gain. When two coupled modes have a specific difference between their loss or gain, a feature termed an exceptional point emerges in the excitations' energy manifold, at which both the eigenfrequencies and eigenmodes of the system coalesce~\cite{el-ganainyNonHermitianPhysicsPT2018, ozdemirParityTimeSymmetry2019}. Exceptional points have intriguing effects on the dynamics of systems due to their topological properties. They have been explored in contexts including optical~\cite{pengParityTimesymmetricWhisperinggallery2014}, microwave~\cite{dietzRabiOscillationsExceptional2007, dopplerDynamicallyEncirclingExceptional2016, partanenExceptionalPointsTunable2019}, optomechanical~\cite{xuTopologicalEnergyTransfer2016, zhangPhononLaserOperating2018}, electronic~\cite{stehmannObservationExceptionalPoints2004, choiObservationAntiPTsymmetricExceptional2018, choiDirectObservationTimeasymmetric2020} and magnonic systems~\cite{zhangObservationExceptionalPoint2017, liuObservationExceptionalPoints2019, zhangExperimentalObservationExceptional2019}, and have been used to control systems including optical microcavities~\cite{jiangCoherentControlChaotic2023}, the lasing modes of a PT-symmetric waveguide~\cite{schumerTopologicalModesLaser2022}, and terahertz pulse generation~\cite{ergoktasTopologicalEngineeringTerahertz2022}. A challenging problem that remains open in all of these scenarios is the fully deterministic and direct manipulation of the systems' loss and gain on timescales relevant to coherent control of excitations. Here we demonstrate the rapid manipulation of the gain and loss balance of excitations of a magnonic hybrid system on durations much shorter than their decay rate, allowing us to exploit non-Hermitian physics for coherent control. By encircling an exceptional point~\cite{yoonTimeasymmetricLoopExceptional2018, feilhauerEncirclingExceptionalPoints2020}, we demonstrate population transfer between coupled magnon-polariton modes, and confirm the distinctive chiral nature of exceptional point encircling. We then study the effect of driving the system directly through an exceptional point, and demonstrate that this allows the coupled system to be prepared in an equal superposition of eigenmodes. We also show that the dynamics of the system at the exceptional point are dependent on its generalised eigenvectors. These results extend the established toolbox of adiabatic transfer techniques with a new versatile approach for coherent state preparation. The highly controllable nature of our hybrid platform provides a new avenue for exploring the intriguing dynamical properties of non-Hermitian systems.
\end{abstract}

\maketitle


Coherent interactions between light and matter are of great fundamental interest, as well as lying at the heart of many applications. The coupling between microwave radiation and the collective excitations of ordered ensembles of spins, termed magnons, has attracted particular attention in recent years. In these systems, the strong coupling regime can be readily reached because of both the confinement of microwaves in the electromagnetic cavity, and the collective enhancement due to the large number of spins present in the material. In these cavity magnonic devices, the hybrid coupled modes are termed magnon polaritons~\cite{PhysRevApplied.2.054002, evertsUltrastrongCouplingMicrowave2020a, lambertIdentificationSpinWave2015, tabuchiHybridizingFerromagneticMagnons2014, zhangStronglyCoupledMagnons2014}. They are a candidate system for quantum information processing due to their long excitation lifetimes and wide ranging frequency tunability~\cite{lachance-quirionHybridQuantumSystems2019, zarerameshtiCavityMagnonics2022}, and have allowed demonstrations of cavity mediated coupling to qubits~\cite{tabuchiCoherentCouplingFerromagnetic2015, lachance-quirionEntanglementbasedSingleshotDetection2020} and other magnon modes~\cite{lambertCavitymediatedCoherentCoupling2016a}, non-reciprocal devices~\cite{wangNonreciprocityUnidirectionalInvisibility2019, zhangBroadbandNonreciprocityEnabled2020} and  non-Hermitian physics.~\cite{zhangObservationExceptionalPoint2017, liuObservationExceptionalPoints2019, zhangExperimentalObservationExceptional2019, caoExceptionalMagneticSensitivity2019, harderCoherentDissipativeCavity2021, hurstNonHermitianPhysicsMagnetic2022, qianNonHermitianControlAbsorption2023, wangEnhancementMagnonicFrequency2024}

An essential requirement to exploit the full potential of hybrid systems is the coherent control of the excitations of the coupled resonances. A number of protocols to achieve this have been demonstrated in non-magnonic platforms by control of frequencies and couplings on the timescale of the coherence times of the excitations.~\cite{vitanovStimulatedRamanAdiabatic2017} Transferring such protocols to magnonic systems is challenging, however, due to the difficulties of generating a rapidly changing magnetic field to tune the frequencies of magnon modes. As an alternative route, Floquet systems have been explored using sinusoidal perturbations of the real part of the frequency in a cavity electromagnonic system~\cite{xuFloquetCavityElectromagnonics2020, qiFloquetGenerationMagnonic2023, yangTheoryFloquetdrivenDissipative2023, zhangNonHermitianShortcutAdiabaticity2022, zhuFloquetengineeringMagnonicNOON2023}. This leaves the full rapid manipulation of complex frequencies necessary for non-Hermitian control in magnonic devices as an outstanding open problem.

In this work, we investigate non-Hermitian control of a hybrid cavity magnonic system comprising two yttrium iron garnet spheres embedded in two coupled active microwave resonators. The Hamiltonian of the system is controllable on timescales as short as $\sim \qty{10}{\nano\second}$ via applied voltage waveforms, allowing us to navigate along arbitrary pathways on the non-Hermitian manifold of the system. We demonstrate on demand transfer of energy from one magnon-polariton mode to another by encircling an EP, and examine the behaviour of the system close to the EP, showing that the dynamics of the generalised eigenbasis must be considered at this point.

\begin{figure*}
\includegraphics{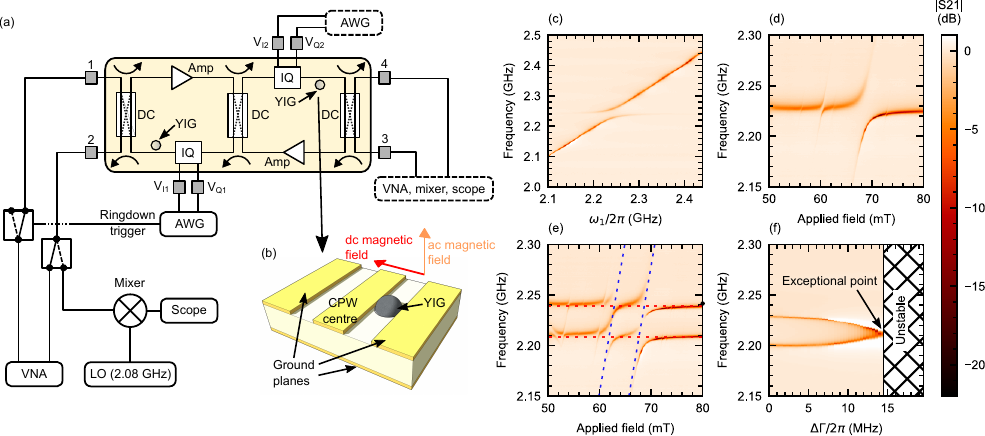}
\caption{\textbf{Experimental configuration and strong coupling.} (a) Schematic of the coupled resonators and readout electronics, showing coupling ports 1 -- 4, amplifiers (Amp), directional couplers (DC), IQ modulators (IQ) and yttrium iron garnet spheres (YIG). (b) The position of the YIG in the coplanar waveguide of the resonator. Also shown is the direction of the field components at the location of the YIG sphere. (c) Feedline transmission \s{21} as a function of waveguide loop resonator 1 frequency and probe frequency, with the frequency of resonator 2 fixed at \qty{2.225}{\giga\hertz} and magnon modes strongly detuned. An avoided crossing betweeen microwave resonances is seen, showing that they are in the strong coupling regime. (d) Transmission as a function of applied magnetic field and probe frequency, with the frequency of resonator 1 fixed at \qty{2.225}{\giga\hertz} and resonator 2 suppressed. The real part of the frequencies of the magnon modes anticross with the microwave mode, demonstrating that the yare strongly coupled. (e) Transmission as a function of applied magnetic field and probe frequency, with the uncoupled frequencies of resonators 1 and 2 fixed at \qty{2.225}{\giga\hertz}, showing the complete mode manifold. Red dotted lines show the frequencies of the microwave photon supermodes, and blue dotted lines show the uncoupled frequencies of the magnon modes. (f) Transmission as a function of damping detuning $\Delta\Gamma$ between the two resonators, with the uncoupled frequencies of resonators 1 and 2 fixed at \qty{2.225}{\giga\hertz} and the applied magnetic field fixed at \qty{68} {\milli\tesla}. An exceptional point is reached at $\Delta\Gamma/2\pi = g/2\pi = \qty{13.5}{\mega\hertz}$.}\label{fig:devcharac}
\end{figure*}

The Hamiltonian for two coupled modes with angular frequencies $\omega_{1(2)}$ and loss rates $\Gamma_{1(2)}$ is
\begin{align}\label{eqn:Ham}
\mathbf{H} = \hbar
\begin{pmatrix}
\omega_1-i\Gamma_1 & g \\
g & \omega_2 -i\Gamma_2
\end{pmatrix},
\end{align}
where $g$ is the strength of the coupling between modes 1 and 2. Negative $\Gamma_{1(2)}$ corresponds to gain in that mode, rather than loss. Strong coupling is reached when the coupling rate is much greater than the loss rate in both of the modes. In many such systems, the loss rates are positive and approximately equal. In this regime the real parts of the eigenvalues of the Hamiltonian exhibit an anticrossing, while their imaginary parts coalesce. Directly at the anti-crossing, one obtains two eigenmodes separated in angular frequency by $2g$, and with loss rates $(\Gamma_1 + \Gamma_2)/2$.  On the other hand, if the difference in the loss rates is large enough such that $|\Gamma_1 - \Gamma_2| > 2g$, an avoided crossing opens between the imaginary parts of the eigenvalues, and the real parts coalesce. The point separating the two regimes, where $|\Gamma_1 - \Gamma_2| = 2g$ and $\omega_1 = \omega_2$, is the exceptional point. Here  the number of distinct eigenvectors and eigenvalues is reduced to one.

The study of non-Hermitian dynamics requires that the complex frequencies of the system's modes can all be fully controlled. This is made possible in our experiment by a novel active microwave resonator (Fig.~\ref{fig:devcharac}), comprising a rectangular loop of \qty{50}{\ohm} coplanar waveguide on low loss PCB, with a footprint of $\qty{20}{\milli\meter} \times \qty{16}{\milli\meter}$ (see Methods). The supported eigenmodes correspond to travelling waves with periodic boundary conditions imposing a $2n\pi$ phase shift around the loop, where $n$ is an integer. An embedded microwave amplifier provides a fixed gain, and an IQ modulator/demodulator allows high bandwidth ($\sim\qty{1}{\giga\hertz}$) quadrature control over the phase and amplitude of the propagating field via control voltages $V_I$ and $V_Q$ at its I and Q inputs. By adjusting the angle $\atantwo(V_Q,V_I)$ the eigenfrequencies can be changed by adding an offset phase to the field. Increasing $\sqrt{V_I^2 + V_Q^2}$ increases the amplitude of the field and decreases the linewidths of the eigenmodes. Setting the voltages to zero effectively disables the resonator by suppressing the travelling wave, while the gain regime can be accessed by sufficiently large values of the control voltages.

Two nominally identical such resonators are coupled together using a stripline directional coupler with a coupling of \qty{-16}{\decibel} (Fig.~\ref{fig:devcharac}(a)). Each resonator is also separately coupled to a \qty{50}{\ohm} feedline with an identical directional coupler, allowing the resonators to be probed by measuring the complex transmission of the feedlines using a vector network analyser (VNA). In Fig.~\ref{fig:devcharac}(c) we show the interaction between two microwave modes, one in each resonator. We slowly sweep the centre frequency of resonator 1 from \qty{2.1}{\giga\hertz} to \qty{2.44}{\giga\hertz} by adjusting the control biases while keeping the frequency of resonator 2 at \qty{2.225}{\giga\hertz} and fixing both mode linewidths to be $\Gamma_{1,2}/2\pi = \qty{2.5}{\mega\hertz}$. By measuring the transmission, we observe an anticrossing between the modes, and find a coupling strength of $g_{12}/2\pi = \qty{15.2(1)}{\mega\hertz} \gg \qty{2.5}{\mega\hertz}$, demonstrating that the two resonator modes are in the strong coupling regime.

An yttrium iron garnet (YIG) sphere of diameter \qty{1}{\milli\metre} is embedded in each resonator by placing it in the gap between the coplanar wave guide's inner and ground plane~\cite{Morris2017} (Fig.~\ref{fig:devcharac}(b)) and a uniform d.c.~magnetic field $H$ is applied in the plane of the waveguide using an electromagnet. YIG is a popular choice for magnonic devices due to its high spin density, and polished YIG spheres support a well understood family of low-loss magnetostatic modes~\cite{PhysRev.105.390, PhysRev.114.739}. The sphere is close to the amplitude maximum of the a.c.~magnetic field generated by the high frequency current in the stripline in order to maximise the coupling between photons and magnons. Furthermore, the field at this point is approximately spatially uniform such that the coupling to the uniform (Kittel) magnetostatic mode is dominant, and the coupling to higher order modes is suppressed~\cite{lambertIdentificationSpinWave2015}.

We probe the interaction between one resonator and the magnons in the co-located sphere by measuring feed line transmission as above, with the second resonator switched off. In Fig.~\ref{fig:devcharac}(e) we show transmission with $\omega_1/2\pi = \qty{2.225}{\giga\hertz}$ and $\Gamma_1/2\pi = \qty{2.5}{\mega\hertz}$, and $H$ swept from \qty{50}{\milli\tesla} to \qty{80}{\milli\tesla} in order to tune the magnon frequency.  We observe an anticrossing between the magnetostatic and microwave modes, demonstrating that the magnon-photon coupling is also in the strong coupling regime, with the hybridised modes being termed magnon-polaritons.  Similar results are seen for Resonator 2 by measuring transmission from port 3 to port 4, and we find values for the coupling strengths of $g_{1,m}/2\pi = \qty{21.3(1)}{\mega\hertz}$ and $g_{2,m}/2\pi = \qty{18.9(1)}{\mega\hertz}$.

We finally characterise the complete magnon-polariton system by again sweeping the applied magnetic field between \qty{50}{\milli\tesla} to \qty{80}{\milli\tesla}, with the uncoupled frequencies of both resonators tuned to \qty{2.225}{\giga\hertz}. We also spatially offset the resonators within the magnetic field, such that one magnon population is detuned by $\sim2g_{12}$. Anticrossings between the magnon-polariton modes are observed (Fig.~\ref{fig:devcharac}(e)). 

With the frequency of the uncoupled photon modes still fixed at \qty{2.225}{\giga\hertz}, we now apply a fixed magnetic field of \qty{68}{\milli\tesla} in the plane of the PCB. Appropriate choices of control biases set the frequencies of two uncoupled magnon-polariton modes (labelled $\ket{1}$ and $\ket{2}$) to be $\omega_1/2\pi = \omega_2/2\pi = \omega_0/2\pi=  \qty{2.2072}{\giga\hertz}$, and their loss rates to be $\Gamma_{1}/2 \pi \approx \Gamma_{2}/2 \pi \approx \Gamma_{0}/2 \pi = \qty{2.5}{\mega\hertz}$. The mode coupling strength is $g/2\pi = \qty{13.5}{\mega\hertz}$, resulting in the eigenstates being symmetric and antisymmetric linear combinations of the uncoupled magnon polariton modes, with frequencies $\omega_l/2 \pi = \qty{2.1937}{\giga\hertz}$ and $\omega_u/2 \pi = \qty{2.2207}{\giga\hertz}$. We label the eigenstates label $\ket{l}$ and $\ket{u}$ respectively.

We now study the behaviour of the coupled magnon-polariton modes as function of the loss rates in the uncoupled modes. We set
\begin{align}
\begin{split}
\Gamma_1 &= \Gamma_0 + \Delta \Gamma,\\
\Gamma_2 &= \Gamma_0 - \Delta \Gamma,
\end{split}
\end{align}
with $2\Delta\Gamma$ being the detuning of the loss rates of between $\ket{1}$ and $\ket{2}$. In Fig.~\ref{fig:devcharac}(f) we show feedline transmission as a function of loss rate detuning. Level attraction between the eigenfrequencies of the coupled magnon--polaritons is observed with increasing detuning, with the levels coalescing at the exceptional point at $\Delta \Gamma/2\pi \approx g/2\pi = \qty{13.5}{\mega\hertz}$. In the region $\Delta \Gamma > g$, the avoided crossing between the imaginary part of the eigenvalues of the coupled modes results in one mode having a large gain, preventing quasistatic measurements.

\begin{figure}
\includegraphics{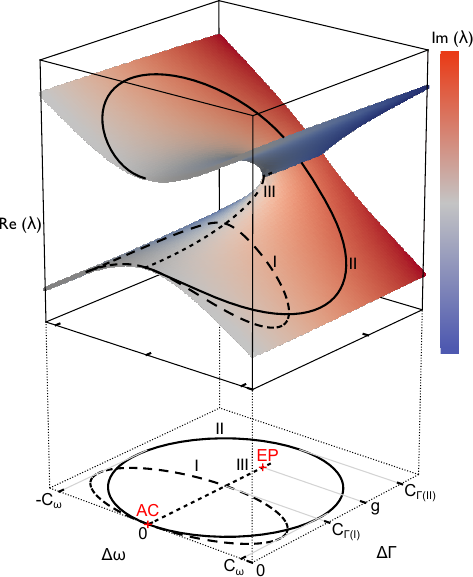}
\caption{\textbf{The theoretical energy landscape for two non-Hermitian coupled resonators}. The surface shows the real parts of the eigenfrequencies ($\textrm{Re}(\lambda$)) of the coupled system as a function of the angular frequency ($\Delta\omega$) and loss ($\Delta\Gamma$) detuning of the resonators from degeneracy, and is coloured according to the imaginary parts of the eigenfrequencies ($\textrm{Im}(\lambda$)). For $\Delta\Gamma < g$ the real part of the eigenvalues exhibits an anticrossing as a function of $\Delta\omega$, and when $\Delta\Gamma > g$ the anticrossing is in the imaginary part. The two regimes are separated by the exceptional point (EP). Also shown are trajectories on the surface and their projection on to the $\{\Delta\omega$, $\Delta\Gamma\}$ plane corresponding to: (I) an ellipse which does not enclose the EP; (II) an ellipse which does enclose the EP; (III) a trajectory from $\Delta\omega= \Delta\Gamma=0$ through the EP and back to the starting point.} \label{fig:Trajectories}
\end{figure}

To further reduce the dimensionality of the parameter space, we now fix
\begin{align}
\begin{split}
\omega_1 &= \omega_0 + \Delta \omega,\\
\omega_2 &= \omega_0 - \Delta \omega,
\end{split}
\end{align}
where $2\Delta\omega/2\pi$ is the frequency detuning between $\ket{1}$ and $\ket{2}$. The resulting calculated energy surface is shown in Fig.~\ref{fig:Trajectories}, with the real part of the eigenvalues plotted as a function of angular frequency detuning $\Delta \omega$ and loss rate detuning $\Delta \Gamma$ of the uncoupled modes. The surface is coloured according to the imaginary part of the eigenvalues, with blue corresponding to $\Gamma>0$ and red to $\Gamma<0$.

\begin{figure}
\includegraphics{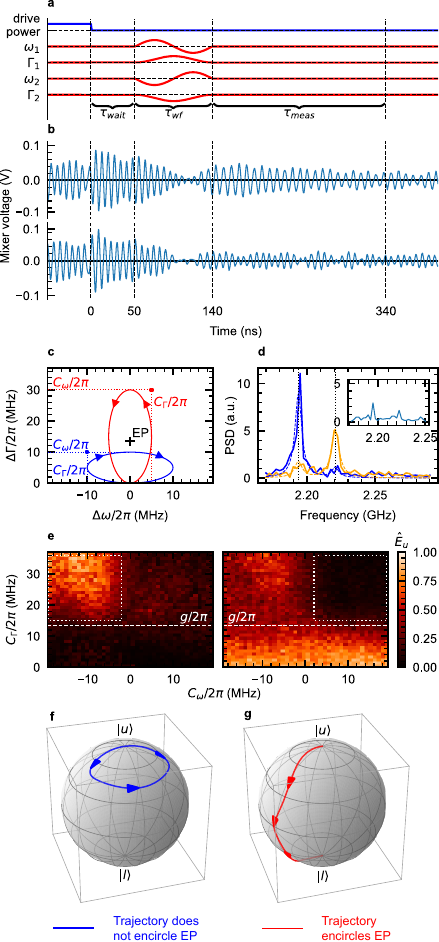}
\caption{\textbf{Population transfer by encircling an exceptional point.} (a) Normalized temporal profile of applied microwave power, and magnon-polariton angular frequencies ($\omega_1$, $\omega_2$) and dampings ($\Gamma_1$, $\Gamma_2$) during the experimental sequence. (b) IF signal for a trajectory encircling the exceptional point on the low loss (upper panel) and high loss (lower panel) surfaces. (c) Example elliptical trajectories (Eq. \ref{eqn:ellipse}) in $\{\Delta\omega, \Delta\Gamma\}$, with  $C_\omega/2\pi = \qty{-10}{\mega\hertz}, C_\Gamma/2\pi = \qty{10}{\mega\hertz}$ (blue, not enclosing EP), $C_\omega/2\pi = \qty{5}{\mega\hertz}, C_\Gamma/2\pi = \qty{30}{\mega\hertz}$ (red, enclosing EP). (d) Power spectra of ringdowns shown in panel (b) during $\tau_\textrm{meas}$. The blue curve correspond to the upper panel and orange to the lower, with fits to a double Lorentzian also shown (dashed curves). Inset: power spectrum at equilibrium, showing excess occupancy of modes above the background. (e) Relative population $\hat{E}_u$ as a function of $C_\omega$ and $C_\Gamma$ for initial excitation of $\ket{l}$ (left) and $\ket{u}$ (right). Population transfer occurs for the parameter ranges marked by the white dotted boxes, where $C_\Gamma>g$ and the trajectory has the correct chirality. (f, g) Path of eigenstates of the system on the Bloch sphere along trajectories in (c). (f) A trajectory which does not encircle the EP returns to its starting point. (g) An EP-encircling trajectory, showing orthogonal start and end points.}\label{fig:EPcircling}
\end{figure}

Our objective now is to carry out coherent manipulations of excitations of $\ket{l}$ and $\ket{u}$ within their lifetime by navigating the coupled system along closed trajectories on the non-Hermitian energy surface. In order to probe the consequences of the trajectory on the excitations, we continuously observe the output waveform at ports 2 and 4, sampled by mixing with a local oscillator at \qty{2.08}{\giga\hertz}. We use the first \qty{200}{\nano\second} of the ringdown following the end of the voltage waveforms to determine the final state of the system. During the ringdown period the voltage in the resonators is described by
\begin{align}
    V(t) = A_l e^{(i\omega_l-\Gamma_l)t} + A_u e^{(i\omega_u-\Gamma_u+\phi)t}
\end{align}
for the corresponding state (up to a global phase)
\begin{align}
    \Psi(t) = A_l \ket{l} + e^{i((\omega_l-\omega_u)t+\phi)} A_u\ket{u},
\end{align}
where $A_l$ and $A_u$ are the absolute amplitudes of $\ket{l}$ and $\ket{u}$ at the beginning of the ringdown, and $\phi$ is their relative phase. The normalised population of $\ket{u}$ is $\hat{E}_u = A_u^2/(A_l^2+A_u^2)$. To determine $A_l$ and $A_u$ we take the power spectrum of the output waveform during the ringdown, summing the contribution from ports 2 and 4. The amplitude of peaks at the frequencies of $\ket{l}$ and $\ket{u}$ are extracted by fitting a double Lorentzian peak to the spectrum (see Methods).

We also observe the equilibrium spectrum in the absence of an applied drive and find peaks above the background at the mode frequencies (Fig.~\ref{fig:EPcircling}(d) inset). These correspond to excess thermal occupancy of the magnon-polariton modes, which we ascribe to the finite noise temperature of the amplifiers embedded in the resonators. This places lower and upper bounds on $\hat{E}_u$, which for our experiments was typically $0.15\lesssim\hat{E}_u\lesssim0.85$.

A fundamental manipulation of the population of a system of normal modes is the transfer of energy from one mode to another. Recent theoretical~\cite{gilaryTimeasymmetricQuantumstateexchangeMechanism2013, uzdinObservabilityAsymmetryAdiabatic2011} and experimental~\cite{dopplerDynamicallyEncirclingExceptional2016, feilhauerEncirclingExceptionalPoints2020, schumerTopologicalModesLaser2022, xuTopologicalEnergyTransfer2016, yoonTimeasymmetricLoopExceptional2018} studies of non-Hermitian systems have investigated and demonstrated such transfer of population by the encircling of an EP. We therefore begin by examining trajectories corresponding to ellipses in $\{\Delta\omega, \Delta\Gamma\}$ (Fig.~\ref{fig:Trajectories}), starting and finishing at the point where loss and frequency detunings are zero ($\Delta\omega = \Delta\Gamma\ =0$). The size of an ellipse determines whether or not it encloses an EP. Furthermore, the trajectory can be traversed in either direction. Voltage waveforms are applied such that
\begin{equation}
\begin{aligned}
    \Delta \omega(t) &= C_\omega\sin{(2\pi (t-t_\textrm{start})/\tau_\textrm{wf})},\\
    \Delta\Gamma(t) &= \tfrac{1}{2}C_{\Gamma}\left(1 - \cos \left(\frac{2\pi (t-t_\textrm{start})}{\tau_\textrm{wf}}\right)\right)
\end{aligned}\label{eqn:ellipse}
\end{equation}
over the time interval between $t_\textrm{start}$ and $t_\textrm{start} + \tau_\textrm{wf}$ with $\tau_\textrm{wf}$ being the duration of the trajectory (Fig.~\ref{fig:EPcircling}(a)). The amplitude of the trajectory is determined by the parameters $C_\omega$ and $C_\Gamma$, representing the maximum deviations from zero detunings for $\omega$ and $\Gamma$ respectively. Changing the sign of $C_\omega$ corresponds to swapping the direction in which the ellipse is traversed. $\tau_\textrm{wf}$ is initially chosen to be \qty{75}{\nano\second}, much shorter than the lifetime of the magnon-polaritons. Example resulting signals for $C_\omega/2\pi = \qty{10.25}{\mega\hertz}$, $C_\Gamma/2\pi = \qty{30}{\mega\hertz}$ (upper panel) and $C_\omega/2\pi = \qty{-10.25}{\mega\hertz}$, $C_\Gamma/2\pi = \qty{30}{\mega\hertz}$ (lower panel) are shown in Fig.~\ref{fig:EPcircling}(b). In both cases $\ket{u}$ was initially populated. 

In Fig.~\ref{fig:EPcircling}(e) we show the normalised population of $\ket{u}$, $\hat{E}_u$, as a function of ellipse size in $\Delta\omega$ and $\Delta\Gamma$ for initial driving of $\ket{l}$ (left panel) and $\ket{u}$ (right panel). When $C_\Gamma<g$ the trajectory does not enclose the EP (I in Fig.~\ref{fig:Trajectories}) and the final state occupancy does not differ significantly from the initial state. However, for trajectories where $C_\Gamma>g$ (II in Fig.~\ref{fig:Trajectories}) the EP is then encircled. The chirality of the encircling is reflected in the measured data, as $\hat{E}_u$ depends on the sense in which the EP is encircled; for an initial population in $\ket{l}$ ($\ket{u}$) $C_\omega$ must be negative (positive) for energy transfer to occur.

A convenient representation of state of a two-level system is given by the Bloch sphere (Figs~\ref{fig:EPcircling}(f) and \ref{fig:EPcircling}(g)), with the poles fixed at $\ket{u}$ and $\ket{l}$. We show the paths of the eigenmodes of the system for the two detuning trajectories in Fig.~\ref{fig:EPcircling}(c). In Fig.~\ref{fig:EPcircling}(f) we show a trajectory for which $C_\Gamma$ is not sufficiently large to encircle the EP, resulting in a closed loop with the state vector starting and terminating at $\ket{u}$. In Fig.~\ref{fig:EPcircling}(g) we show an EP-enclosing trajectory, forming a quasi-adiabatic path from $\ket{u}$ to $\ket{l}$.  The size of the detuning ellipses gives the precise route across the Bloch sphere, but the final state is independent of the details of the trajectory; EP-enclosing trajectories therefore result in robust transfer of population from $\ket{u}$ to $\ket{l}$.

These results demonstrate on-demand control over the non-Hermitian dynamics of the magnon-polariton system: energy can be switched between modes by dynamically encircling an EP~\cite{feilhauerEncirclingExceptionalPoints2020, nasariObservationChiralState2022}, with the final state independent of the exact trajectory. The trajectories exhibit a chiral time asymmetry -- for each initial state, one of the encircling directions leads to an energy transfer while the other encircling direction closely returns the state vector back to its initial configuration. For trajectories which do not come near to encircling the EP, energy is not transferred. (We note that deviations from this behavior have previously been discussed.~\cite{hassanChiralStateConversion2017, nasariObservationChiralState2022})

Rather than robustly transferring a given state to another one, it is often desirable to instead prepare the system in a superposition of states, in which energy is split between the two modes. To realise this goal, we will therefore investigate a strategy in which the EP is approached directly along the real anticrossing (trajectory III in Fig.~\ref{fig:Trajectories}). At the EP, the Hamiltonian cannot be diagonalised, and has only a single eigenvector. Because the two eigenvectors coalesce, it has been suggested~\cite{znojilPassageExceptionalPoint2020} that the defective Hamiltonian erases any history of the trajectory when it passes through the EP. This would result in the desired deterministic preparation of a specific superposition of states, regardless of initial state.

\begin{figure*}
\includegraphics{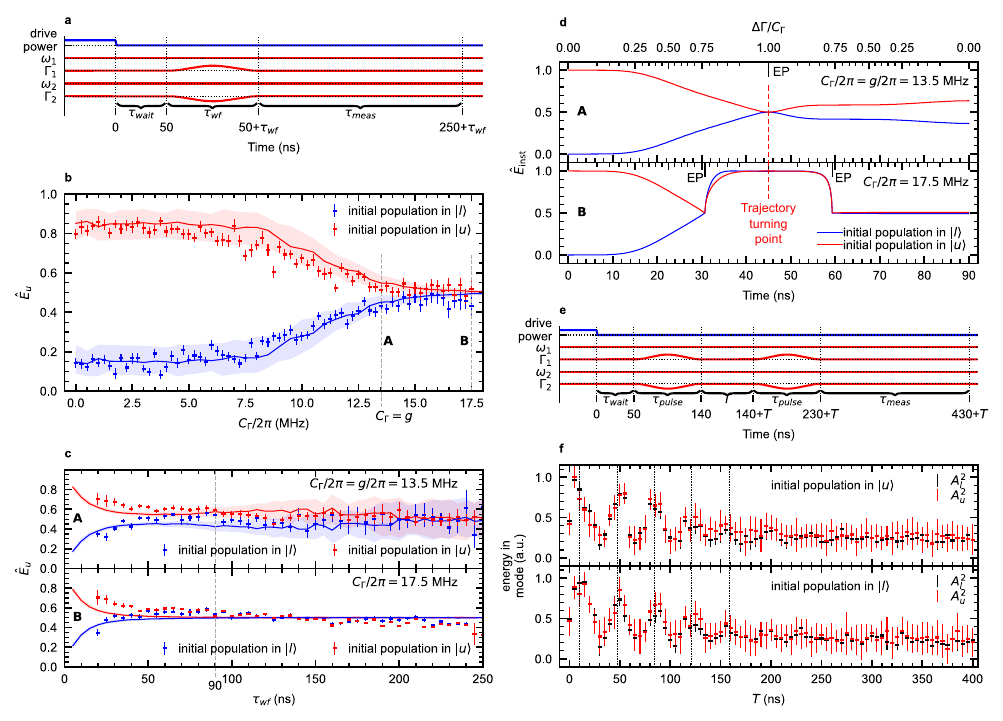}
\caption{\textbf{Equalising state populations by traversing beyond and back through an EP.} (a) Temporal profile of applied microwave power, and resonator angular frequencies ($\omega_1$, $\omega_2$) and loss rates ($\Gamma_1$, $\Gamma_2$) for trajectory III in Fig.~\ref{fig:Trajectories}. (b) Final state population of coupled magnon-polaritons $\hat{E}_u$ as a function of $C_\Gamma$. Error bars show experimental data, and the solid line (shaded region) the mean (standard deviation) of the stochastic model described in the text. $\hat{E}_u$ tends towards 0.5 as $C_\Gamma$ is increased beyond the EP (located at vertical line \textbf{A} at which $C_\Gamma=g$). (c, upper panel) $\hat{E}_u$ as a function of trajectory timespan for $C_\Gamma/2\pi = g/2\pi = \qty{13.5}{\mega\hertz}$ (touching the EP). Populations of $\ket{u}$ and $\ket{l}$ are not equal, even for long duration trajectories. (c, lower panel) $\hat{E}_u$ as a function of trajectory timespan for $C_\Gamma/2\pi = \qty{17.5}{\mega\hertz}$ (vertical line \textbf{B} on panel (b)). Populations of $\ket{u}$ and $\ket{l}$ are now equalised for durations longer than $\tau_\textrm{wf}\approx\qty{80}{\nano\second}$. The effect of thermal noise is much smaller due to the higher absolute mode populations. (d) Calculated instantaneous eigenvector populations during trajectories with $C_\Gamma/2\pi = \qty{13.5}{\mega\hertz}$ (upper panel) and $C_\Gamma/2\pi = \qty{17.5}{\mega\hertz}$ (lower panel), and for $\tau_\textrm{wf} = \qty{90}{\nano\second}$ (dotted line in panel (c)). Preparation of an equal admixture of $\ket{l}$ and $\ket{u}$ relies on the contrast between loss and gain states between \qty{30}{\nano\second} and \qty{60}{\nano\second} for $C_\Gamma/2\pi = \qty{17.5}{\mega\hertz}$. (e) Temporal profile of applied microwave power, and resonator angular frequencies and loss rates for two consecutive trajectories towards the EP. (f) Energies in $\ket{l}$ and $\ket{u}$ after consecutive trajectories beyond the EP ($C_\Gamma/2\pi = \qty{17.5}{\mega\hertz}$) as a function of time $T$ between consecutive trajectories, for initial population in $\ket{u}$ (upper panel) and $\ket{l}$ (lower panel). Mode energies oscillate in-phase regardless of the initially populated state (dotted lines), demonstrating the independence of final state on initial state.}\label{fig:DirectEP}
\end{figure*}

To test this idea, we investigate trajectories that lie along the real anticrossing (trajectory III in Fig.~\ref{fig:Trajectories}) such that
\begin{equation}
\label{eqn:direct}
\begin{aligned}
    \Delta \omega &= 0,\\
    \Delta\Gamma(t) &= \tfrac{1}{2}C_\Gamma\left(1 - \cos \left(\frac{2\pi (t-t_\textrm{start})}{\tau_\textrm{wf}}\right)\right).
\end{aligned}
\end{equation}
The resulting eigenvector populations are measured as before, initially with $\tau_\textrm{wf} = \qty{90}{\nano\second}$ and $0 \leq C_\Gamma/2\pi \leq \qty{17.5}{\mega\hertz}$. In Fig.~\ref{fig:DirectEP}(b) we plot the normalised population of $\ket{u}$ as a function of increasing trajectory length $C_\Gamma$ for an initial population in $\ket{l}$ (blue) and $\ket{u}$ (red). For small excursions towards the EP, there is little transfer of energy between $\ket{l}$ and $\ket{u}$, and the final state is dominated by the initially excited eigenstate. On the other hand, as the maximum loss detuning approaches the value of the coupling strength $g$ (i.e., the location of the EP), the proportion of energy transferred between the modes increases. However, we find that $\hat{E}_u$ does not reach $0.5$ for $C_\Gamma/2\pi = g/2\pi = \qty{13.5}{\mega\hertz}$, despite these trajectories touching the EP. The populations of $\ket{l}$ and $\ket{u}$ are therefore not equalised at the EP, but only when extending the trajectory beyond the EP to the value $C_\Gamma/2\pi \approx \qty{17}{\mega\hertz}$.

To understand this interesting behaviour, we model the temporal evolution of the state vector numerically under the time dependent Hamiltonian of the system~\cite{feilhauerEncirclingExceptionalPoints2020}. In this simulation, white Gaussian noise is added that represents the effect of the amplifier noise, and we calculate the overlaps of the final state with $\ket{u}$ and $\ket{l}$. The evolution is parameterised by $g/2\pi=\qty{13.5}{\mega\hertz}, \Gamma_0/2\pi = \qty{2.5}{\mega\hertz}$, and no free parameters other than the noise amplitude are used (which is determined from fitting the output state of the model in the absence of applied pulses to the measured equilibrium occupation of the modes.) The stochastic simulation is run 1000 times, and the mean and standard deviation are shown in Fig.~\ref{fig:DirectEP}(b), showing excellent agreement with experimental data.

Final populations are affected both by the amount of time spent near the EP and the total time $\tau_\textrm{wf}$ over which the trajectories are traversed. We investigate this dependence by measuring the normalised population of $\ket{u}$ as a function of the duration of the trajectory, varying $\tau_\textrm{wf}$ from \qty{20}{\nano\second} to \qty{250}{\nano\second}. Results for a trajectory that just reaches the EP ($C_\Gamma/2\pi = \qty{13.5}{\mega\hertz}$, line \textbf{A} in panel (b)) and for one that goes beyond the EP ($C_\Gamma/2\pi = \qty{17.5}{\mega\hertz}$, line \textbf{B} in panel (b))) are shown in Fig.~\ref{fig:DirectEP}(c).

In both the simulations and experiments we find that the desired equalisation of populations does not occur at the critical value $C_\Gamma/2\pi = \qty{13.5}{\mega\hertz}$, even for long duration trajectories corresponding to large values of $\tau_\textrm{wf}$. Contrary to our naive expectation, reaching the EP therefore does not equalise the population of the eigenstates. Our theoretical analysis shows that the reason for this behaviour lies in the fact that the two eigenvectors that merge at the EP fail to provide a full description of the state of the system. To restore a complete basis, the generalised eigenvectors of the Hamiltonian need to be included (see Methods), which are populated at the EP regardless of the initial population of the modes. With the generalised eigenvector preventing an erasure of the system's history before reaching the EP, an excursion to the EP alone is therefore not an effective state preparation technique.

The normalised populations of $\ket{l}$ and $\ket{u}$ do, however, reach the desired value of $0.5$ at $C_\Gamma/2\pi = \qty{17.5}{\mega\hertz}$ and for $\tau_\textrm{wf}\gtrsim\qty{80}{\nano\second}$, irrespective of the initial population. To gain further insight into this behaviour, we theoretically study the evolution of the state vector $\ket{\sigma}$ of the system \emph{during} the trajectory by noise-free numeric evolution of the Hamiltonian. In Fig.~\ref{fig:DirectEP}(d) we plot the relative population of the \emph{instantaneous} eigenvectors $\ket{v_1}$ and $\ket{v_2}$ as a function of time during the trajectory, such that $\hat{E}_\textrm{inst} = \lvert\braket{\hat{v}_1|\sigma}\rvert^2/(\lvert\braket{\hat{v}_1|\sigma}\rvert^2+\lvert\braket{\hat{v}_2|\sigma}\rvert^2)$, where the hat denotes the biorthogonal left-eigenvector. This is again plotted for $C_\Gamma/2\pi = g/2\pi = \qty{13.5}{\mega\hertz}$ and $C_\Gamma/2\pi = \qty{17.5}{\mega\hertz}$, and $\tau_\mathrm{wf} = \qty{90}{\nano\second}$. For $C_\Gamma/2\pi = \qty{13.5}{\mega\hertz}$, the EP is reached only momentarily in the middle of the trajectory, which is insufficient to equalise the population of the instantaneous eigenvectors. For $C_\Gamma/2\pi = \qty{17.5}{\mega\hertz}$, the EP is crossed twice, at $t\approx\qty{30}{\nano\second}$ and $t\approx\qty{60}{\nano\second}$. The first EP crossing parallelizes the eigenvectors and ensures a non-zero population of both instantaneous eigenvectors. In the regime beyond the EP where $\Delta\Gamma>g$, the real part of the eigenfrequencies are equal, but the imaginary parts deviate from each other. This results in one eigenstate having gain, and the other loss. As a consequence, the gain eigenstate grows in amplitude and the loss eigenstate is suppressed, resulting in the population being driven predominantly to the gain state. Traversing back through the EP to the zero detuning point projects the state into an equal admixture of $\ket{l}$ and $\ket{u}$. This leads to consistent preparation of a state for which $A_l = A_u$, as the trajectories along surfaces of equal gain/loss rate freeze the population ratio at 0.5.

While we have confirmed that trajectories beyond the EP can lead to equal populations in $\ket{l}$ and $\ket{u}$, this does not demonstrate that the process is coherent and results in a single final state, as the phase $\phi$ between $\ket{l}$ and $\ket{u}$ is not determined by the above experiments. To probe the coherence of this process, we measure the amplitudes of $\ket{l}$ and $\ket{u}$ after two consecutive identical trajectories through the EP with $C_\Gamma/2\pi = \qty{17.5}{\mega\hertz}$, separated by time $T$ (Fig.~\ref{fig:DirectEP}(e)). The first trajectory prepares the system in an equal superposition of $\ket{l}$ and $\ket{u}$ according to the process described above. The state vector then precesses around the equator of the Bloch sphere during time $T$, as a dynamical phase between $\ket{l}$ and $\ket{u}$ accumulates. While the state of the system is always $\hat{E}_u=0.5$ at the end of the second trajectory, the \textit{magnitudes} of the populations are dependent on the dynamical phase, and therefore on $T$, due to the two states that have been equally populated by the first trajectory undergoing either constructive or destructive interference, in loose analogy to Ramsey interferometry.

We measure final excitation amplitudes for an initial population of $\ket{l}$ (Fig.~\ref{fig:DirectEP}(f), upper panel) and $\ket{u}$ (Fig.~\ref{fig:DirectEP}(f), lower panel). Oscillations of the amplitudes in $T$ are observed, with period $\qty{37}{\nano\second} \approx 2\pi/2g$. For initial population of either mode, the amplitudes of the two final states are in phase and equal (red and black data in each panel). We therefore obtain an equal admixture of states regardless of the phase acquired during $T$. This also demonstrates that the total amplitude of the states is sensitive to the phase between $\ket{l}$ and $\ket{u}$. Furthermore, the amplitudes of the final states are also in phase between initial population of $\ket{l}$ and $\ket{u}$ (comparing upper and lower panels of Fig.~\ref{fig:DirectEP}(f)), showing that the initial state does not affect the final state up to a global amplitude.

In conclusion, we have demonstrated coherent control of magnon-polaritons by coupling magnetostatic modes to agile microwave cavities in which both frequency and gain can be controlled. In particular, we use frequency-gain trajectories encircling the EP of the Riemann sheet of the complex eigenfrequencies of the system to deterministically transfer excitations between magnon-polariton modes. By driving the system to the EP, where both eigenvectors and eigenvalues coalesce, we study the dynamics of its generalised eigensystem, and by following a trajectory going beyond the EP, we demonstrate deterministic preparation of the system in a superposition of magnon-polariton states. 

As well as being a powerful tool to investigate non-Hermitian physics, our platform offers a way to manipulate hybridised states in a variety of systems. Due to the high bandwidth of the tunability of the cavity, the magnon-polaritons can be tuned at rates approaching their frequency, which could make it a useful platform for exploration of non-equilibrium physics in hybrid systems beyond the rotating wave approximation.

\section*{Methods}

\textbf{Microwave resonator} -- The resonator was fabricated on low loss Rogers Material RO4350B, with a substrate thickness of \qty{0.762}{\milli\metre}. Stripline backwards wave directional couplers were all Minicircuits SCBD-16-63HP+. Amplification was provided by Minicircuits YSF-322+ low-noise amplifiers, and IQ modulators were Marki Microwave MMIQ-0205HSM-2. Polished YIG spheres were provided by www.ferrisphere.com.
\\

\textbf{Measurement protocol} -- Our experiments proceed as follows: i) The control voltages and magnetic field are set so that eigenmodes of the system are $\ket{l}$ and $\ket{u}$. ii) Either $\ket{l}$ or $\ket{u}$ is populated by applying a long ($>\qty{10}{\micro\second}$) CW drive ($\textrm{power} = \qty{-35}{\dBm}$) at its resonant frequency via port 1. iii) The drive is switched off using a microwave switch with a switching time $\sim\qty{20}{\nano\second}$. iv) Waveforms are applied to the control voltages, such that the system follows a particular path in \{$\Delta \omega$, $\Delta \Gamma$\} over a time period $\sim\qty{100}{\nano\second}$. Trajectories are always designed such that $\ket{l}$ and $\ket{u}$ are eigenmodes at both beginning and end of the path.  v) The resulting excitations are allowed to ring down.

The output waveform at ports 2 and 4 is amplified by \qty{64}{\dB} (2 $\times$ ABL0600-01-3240), and mixed with a local oscillator at \qty{2.08}{\giga\hertz} (Minicircuits ZX05-63LH-S+). The resulting intermediate frequencies (IF) are sampled at a rate of \qty{2.5}{\giga\sample\per\second}. We use the first \qty{200}{\nano\second} of the ringdown following the end of the applied $I$, $Q$ voltage waveforms to determine the final state of the system. For each experimental data point we average over ten such ringdowns. The frequency variation of amplification chain is carefully calibrated by measuring the S-parameters of both individual sections of the chain, and the entire network.\\
 
\textbf{Generalised eigenbasis at exceptional point} -- At the EP the Hamiltonian is
\begin{align}\label{eqn:Ham2}
\mathbf{H} = \hbar
\begin{pmatrix}
\omega_0-i g & g \\
g & \omega_0 +i g
\end{pmatrix},
\end{align}
with a single eigenvector $(-i,1)$. This clearly does not span the space of the state vector, and a full description of the state of the system is not possible. However, a complete basis is restored by the generalised eigenbasis $\{\mathbf{v}_1, \mathbf{v}_2\}$, such that $(\mathbf{H}-\lambda \mathbf{I})^n\mathbf{v}_n = 0$. This leads to $\mathbf{v}_1 = (-i,1), \mathbf{v}_2 = (i,1)$. Modelling the dynamics of these vectors allows the time evolution of the system to be calculated.

\section*{Acknowledgments}

NJL would like to acknowledge discussions with Associate Professor Jonathan Squire.

\section*{Author Contributions}

N.J.L.~conceptualised the experiment and performed implementation and measurement. The manuscript was written by N.J.L.~with input from all authors. A.S.~and S.R.~carried out theoretical calculations and provided conceptual input. The work was supervised by J.J.L., S.R.~and H.G.L.S.

\section*{Author Information}
The authors declare no competing financial interests. Correspondence and requests for materials should be addressed to N.J.L. (nicholas.lambert@otago.ac.nz).

\bibliography{MagnonEPcircling}

@article{caoExceptionalMagneticSensitivity2019,
  title = {Exceptional Magnetic Sensitivity of {{PT-symmetric}} Cavity Magnon Polaritons},
  author = {Cao, Yunshan and Yan, Peng},
  year = {2019},
  month = jun,
  journal = {Physical Review B},
  volume = {99},
  number = {21},
  pages = {214415},
  publisher = {American Physical Society},
  doi = {10.1103/PhysRevB.99.214415},
  url = {https://link.aps.org/doi/10.1103/PhysRevB.99.214415},
  urldate = {2023-12-06}
}

@article{choiDirectObservationTimeasymmetric2020,
  title = {Direct Observation of Time-Asymmetric Breakdown of the Standard Adiabaticity around an Exceptional Point},
  author = {Choi, Youngsun and Yoon, Jae Woong and Hong, Jong Kyun and Ryu, Yeonghwa and Song, Seok Ho},
  year = {2020},
  month = aug,
  journal = {Communications Physics},
  volume = {3},
  number = {1},
  pages = {1--7},
  publisher = {Nature Publishing Group},
  issn = {2399-3650},
  doi = {10.1038/s42005-020-00409-y},
  url = {https://www.nature.com/articles/s42005-020-00409-y},
  urldate = {2021-03-29},
  copyright = {2020 The Author(s)},
  langid = {english}
}

@article{choiObservationAntiPTsymmetricExceptional2018,
  title = {Observation of an Anti-{{PT-symmetric}} Exceptional Point and Energy-Difference Conserving Dynamics in Electrical Circuit Resonators},
  author = {Choi, Youngsun and Hahn, Choloong and Yoon, Jae Woong and Song, Seok Ho},
  year = {2018},
  month = jun,
  journal = {Nature Communications},
  volume = {9},
  number = {1},
  pages = {2182},
  publisher = {Nature Publishing Group},
  issn = {2041-1723},
  doi = {10.1038/s41467-018-04690-y},
  url = {https://www.nature.com/articles/s41467-018-04690-y},
  urldate = {2021-03-29},
  copyright = {2018 The Author(s)},
  langid = {english}
}

@article{dietzRabiOscillationsExceptional2007,
  title = {Rabi Oscillations at Exceptional Points in Microwave Billiards},
  author = {Dietz, B. and Friedrich, T. and Metz, J. and {Miski-Oglu}, M. and Richter, A. and Sch{\"a}fer, F. and Stafford, C. A.},
  year = {2007},
  month = feb,
  journal = {Physical Review E},
  volume = {75},
  number = {2},
  pages = {027201},
  publisher = {American Physical Society},
  doi = {10.1103/PhysRevE.75.027201},
  url = {https://link.aps.org/doi/10.1103/PhysRevE.75.027201},
  urldate = {2023-05-15}
}

@article{dopplerDynamicallyEncirclingExceptional2016,
  title = {Dynamically Encircling an Exceptional Point for Asymmetric Mode Switching},
  author = {Doppler, J{\"o}rg and Mailybaev, Alexei A. and B{\"o}hm, Julian and Kuhl, Ulrich and Girschik, Adrian and Libisch, Florian and Milburn, Thomas J. and Rabl, Peter and Moiseyev, Nimrod and Rotter, Stefan},
  year = {2016},
  month = sep,
  journal = {Nature},
  volume = {537},
  number = {7618},
  pages = {76--79},
  publisher = {Nature Publishing Group},
  issn = {1476-4687},
  doi = {10.1038/nature18605},
  url = {https://www.nature.com/articles/nature18605},
  urldate = {2023-01-26},
  copyright = {2016 Macmillan Publishers Limited, part of Springer Nature. All rights reserved.},
  langid = {english}
}

@article{el-ganainyNonHermitianPhysicsPT2018,
  title = {Non-{{Hermitian}} Physics and {{PT}} Symmetry},
  author = {{El-Ganainy}, Ramy and Makris, Konstantinos G. and Khajavikhan, Mercedeh and Musslimani, Ziad H. and Rotter, Stefan and Christodoulides, Demetrios N.},
  year = {2018},
  month = jan,
  journal = {Nature Physics},
  volume = {14},
  number = {1},
  pages = {11--19},
  publisher = {Nature Publishing Group},
  issn = {1745-2481},
  doi = {10.1038/nphys4323},
  url = {https://www.nature.com/articles/nphys4323},
  urldate = {2023-01-26},
  copyright = {2017 Nature Publishing Group, a division of Macmillan Publishers Limited. All Rights Reserved.},
  langid = {english}
}

@article{ergoktasTopologicalEngineeringTerahertz2022,
  title = {Topological Engineering of Terahertz Light Using Electrically Tunable Exceptional Point Singularities},
  author = {Ergoktas, M. Said and Soleymani, Sina and Kakenov, Nurbek and Wang, Kaiyuan and Smith, Thomas B. and Bakan, Gokhan and Balci, Sinan and Principi, Alessandro and Novoselov, Kostya S. and Ozdemir, Sahin K. and Kocabas, Coskun},
  year = {2022},
  month = apr,
  journal = {Science},
  volume = {376},
  number = {6589},
  pages = {184--188},
  publisher = {American Association for the Advancement of Science},
  doi = {10.1126/science.abn6528},
  url = {https://www.science.org/doi/10.1126/science.abn6528},
  urldate = {2024-01-29}
}

@article{evertsUltrastrongCouplingMicrowave2020a,
  title = {Ultrastrong Coupling between a Microwave Resonator and Antiferromagnetic Resonances of Rare-Earth Ion Spins},
  author = {Everts, Jonathan R. and King, Gavin GG and Lambert, Nicholas J. and Kocsis, Sacha and Rogge, Sven and Longdell, Jevon J.},
  year = {2020},
  journal = {Physical Review B},
  volume = {101},
  number = {21},
  pages = {214414},
  publisher = {APS}
}

@article{feilhauerEncirclingExceptionalPoints2020,
  title = {Encircling Exceptional Points as a Non-{{Hermitian}} Extension of Rapid Adiabatic Passage},
  author = {Feilhauer, J. and Schumer, A. and Doppler, J. and Mailybaev, A. A. and B{\"o}hm, J. and Kuhl, U. and Moiseyev, N. and Rotter, S.},
  year = {2020},
  month = oct,
  journal = {Physical Review A},
  volume = {102},
  number = {4},
  pages = {040201},
  publisher = {American Physical Society},
  doi = {10.1103/PhysRevA.102.040201},
  url = {https://link.aps.org/doi/10.1103/PhysRevA.102.040201},
  urldate = {2022-08-01}
}

@article{gilaryTimeasymmetricQuantumstateexchangeMechanism2013,
  title = {Time-Asymmetric Quantum-State-Exchange Mechanism},
  author = {Gilary, Ido and Mailybaev, Alexei A. and Moiseyev, Nimrod},
  year = {2013},
  month = jul,
  journal = {Physical Review A},
  volume = {88},
  number = {1},
  pages = {010102},
  publisher = {American Physical Society},
  doi = {10.1103/PhysRevA.88.010102},
  url = {https://link.aps.org/doi/10.1103/PhysRevA.88.010102},
  urldate = {2024-05-18}
}

@article{harderCoherentDissipativeCavity2021,
  title = {Coherent and Dissipative Cavity Magnonics},
  author = {Harder, M. and Yao, B. M. and Gui, Y. S. and Hu, C.-M.},
  year = {2021},
  month = may,
  journal = {Journal of Applied Physics},
  volume = {129},
  number = {20},
  pages = {201101},
  issn = {0021-8979},
  doi = {10.1063/5.0046202},
  url = {https://doi.org/10.1063/5.0046202},
  urldate = {2023-11-29}
}

@article{hassanChiralStateConversion2017,
  title = {Chiral State Conversion without Encircling an Exceptional Point},
  author = {Hassan, Absar U. and Galmiche, Gisela L. and Harari, Gal and LiKamWa, Patrick and Khajavikhan, Mercedeh and Segev, Mordechai and Christodoulides, Demetrios N.},
  year = {2017},
  month = nov,
  journal = {Physical Review A},
  volume = {96},
  number = {5},
  pages = {052129},
  publisher = {American Physical Society},
  doi = {10.1103/PhysRevA.96.052129},
  url = {https://link.aps.org/doi/10.1103/PhysRevA.96.052129},
  urldate = {2024-06-19}
}

@article{hurstNonHermitianPhysicsMagnetic2022,
  title = {Non-{{Hermitian}} Physics in Magnetic Systems},
  author = {Hurst, Hilary M. and Flebus, Benedetta},
  year = {2022},
  month = dec,
  journal = {Journal of Applied Physics},
  volume = {132},
  number = {22},
  pages = {220902},
  issn = {0021-8979},
  doi = {10.1063/5.0124841},
  url = {https://doi.org/10.1063/5.0124841},
  urldate = {2023-11-29}
}

@article{jiangCoherentControlChaotic2023,
  title = {Coherent Control of Chaotic Optical Microcavity with Reflectionless Scattering Modes},
  author = {Jiang, Xuefeng and Yin, Shixiong and Li, Huanan and Quan, Jiamin and Goh, Heedong and Cotrufo, Michele and Kullig, Julius and Wiersig, Jan and Al{\`u}, Andrea},
  year = {2023},
  month = nov,
  journal = {Nature Physics},
  pages = {1--7},
  publisher = {Nature Publishing Group},
  issn = {1745-2481},
  doi = {10.1038/s41567-023-02242-w},
  url = {https://www.nature.com/articles/s41567-023-02242-w},
  urldate = {2023-11-29},
  copyright = {2023 The Author(s), under exclusive licence to Springer Nature Limited},
  langid = {english}
}

@article{lachance-quirionEntanglementbasedSingleshotDetection2020,
  title = {Entanglement-Based Single-Shot Detection of a Single Magnon with a Superconducting Qubit},
  author = {{Lachance-Quirion}, Dany and Wolski, Samuel Piotr and Tabuchi, Yutaka and Kono, Shingo and Usami, Koji and Nakamura, Yasunobu},
  year = {2020},
  month = jan,
  journal = {Science},
  volume = {367},
  number = {6476},
  pages = {425--428},
  publisher = {American Association for the Advancement of Science},
  doi = {10.1126/science.aaz9236},
  url = {https://www.science.org/doi/10.1126/science.aaz9236},
  urldate = {2024-05-16}
}

@article{lachance-quirionHybridQuantumSystems2019,
  title = {Hybrid Quantum Systems Based on Magnonics},
  author = {{Lachance-Quirion}, Dany and Tabuchi, Yutaka and Gloppe, Arnaud and Usami, Koji and Nakamura, Yasunobu},
  year = {2019},
  month = jun,
  journal = {Applied Physics Express},
  volume = {12},
  number = {7},
  pages = {070101},
  publisher = {IOP Publishing},
  issn = {1882-0786},
  doi = {10.7567/1882-0786/ab248d},
  url = {https://doi.org/10.7567/1882-0786/ab248d},
  urldate = {2021-08-19},
  langid = {english}
}

@article{lambertCavitymediatedCoherentCoupling2016a,
  title = {Cavity-Mediated Coherent Coupling of Magnetic Moments},
  author = {Lambert, N. J. and Haigh, J. A. and Langenfeld, S. and Doherty, A. C. and Ferguson, A. J.},
  year = {2016},
  month = feb,
  journal = {Physical Review A},
  volume = {93},
  number = {2},
  pages = {021803},
  publisher = {American Physical Society},
  doi = {10.1103/PhysRevA.93.021803},
  url = {https://link.aps.org/doi/10.1103/PhysRevA.93.021803},
  urldate = {2023-01-25}
}

@article{lambertIdentificationSpinWave2015,
  title = {Identification of Spin Wave Modes in Yttrium Iron Garnet Strongly Coupled to a Co-Axial Cavity},
  author = {Lambert, N. J. and Haigh, J. A. and Ferguson, A. J.},
  year = {2015},
  month = feb,
  journal = {Journal of Applied Physics},
  volume = {117},
  number = {5},
  pages = {053910},
  publisher = {American Institute of Physics},
  issn = {0021-8979},
  doi = {10.1063/1.4907694},
  url = {https://aip.scitation.org/doi/full/10.1063/1.4907694},
  urldate = {2023-01-25}
}

@article{liuObservationExceptionalPoints2019,
  title = {Observation of Exceptional Points in Magnonic Parity-Time Symmetry Devices},
  author = {Liu, Haoliang and Sun, Dali and Zhang, Chuang and Groesbeck, Matthew and Mclaughlin, Ryan and Vardeny, Z. Valy},
  year = {2019},
  month = nov,
  journal = {Science Advances},
  volume = {5},
  number = {11},
  pages = {eaax9144},
  publisher = {American Association for the Advancement of Science},
  issn = {2375-2548},
  doi = {10.1126/sciadv.aax9144},
  url = {https://advances.sciencemag.org/content/5/11/eaax9144},
  urldate = {2021-03-29},
  chapter = {Research Article},
  copyright = {Copyright {\copyright} 2019 The Authors, some rights reserved; exclusive licensee American Association for the Advancement of Science. No claim to original U.S. Government Works. Distributed under a Creative Commons Attribution NonCommercial License 4.0 (CC BY-NC).. This is an open-access article distributed under the terms of the Creative Commons Attribution-NonCommercial license, which permits use, distribution, and reproduction in any medium, so long as the resultant use is not for commercial advantage and provided the original work is properly cited.},
  langid = {english}
}

@article{Morris2017,
  title = {Strong Coupling of Magnons in a {{YIG}} Sphere to Photons in a Planar Superconducting Resonator in the Quantum Limit},
  author = {Morris, R. G. E. and {van Loo}, A. F. and Kosen, S. and Karenowska, A. D.},
  year = {2017},
  month = sep,
  journal = {Scientific Reports},
  volume = {7},
  number = {1},
  pages = {11511},
  issn = {2045-2322},
  doi = {10.1038/s41598-017-11835-4},
  url = {https://doi.org/10.1038/s41598-017-11835-4},
  refid = {Morris2017}
}

@article{nasariObservationChiralState2022,
  title = {Observation of Chiral State Transfer without Encircling an Exceptional Point},
  author = {Nasari, Hadiseh and {Lopez-Galmiche}, Gisela and {Lopez-Aviles}, Helena E. and Schumer, Alexander and Hassan, Absar U. and Zhong, Qi and Rotter, Stefan and LiKamWa, Patrick and Christodoulides, Demetrios N. and Khajavikhan, Mercedeh},
  year = {2022},
  month = may,
  journal = {Nature},
  volume = {605},
  number = {7909},
  pages = {256--261},
  publisher = {Nature Publishing Group},
  issn = {1476-4687},
  doi = {10.1038/s41586-022-04542-2},
  url = {https://www.nature.com/articles/s41586-022-04542-2},
  urldate = {2023-02-09},
  copyright = {2022 The Author(s), under exclusive licence to Springer Nature Limited},
  langid = {english}
}

@article{ozdemirParityTimeSymmetry2019,
  title = {Parity--Time Symmetry and Exceptional Points in Photonics},
  author = {{\"O}zdemir, {\c S} K. and Rotter, S. and Nori, F. and Yang, L.},
  year = {2019},
  month = aug,
  journal = {Nature Materials},
  volume = {18},
  number = {8},
  pages = {783--798},
  publisher = {Nature Publishing Group},
  issn = {1476-4660},
  doi = {10.1038/s41563-019-0304-9},
  url = {https://www.nature.com/articles/s41563-019-0304-9},
  urldate = {2023-01-26},
  copyright = {2019 The Author(s), under exclusive licence to Springer Nature Limited},
  langid = {english}
}

@article{partanenExceptionalPointsTunable2019,
  title = {Exceptional Points in Tunable Superconducting Resonators},
  author = {Partanen, Matti and Goetz, Jan and Tan, Kuan Yen and Kohvakka, Kassius and Sevriuk, Vasilii and Lake, Russell E. and Kokkoniemi, Roope and Ikonen, Joni and Hazra, Dibyendu and M{\"a}kinen, Akseli and Hyypp{\"a}, Eric and Gr{\"o}nberg, Leif and Vesterinen, Visa and Silveri, Matti and M{\"o}tt{\"o}nen, Mikko},
  year = {2019},
  month = oct,
  journal = {Physical Review B},
  volume = {100},
  number = {13},
  pages = {134505},
  publisher = {American Physical Society},
  doi = {10.1103/PhysRevB.100.134505},
  url = {https://link.aps.org/doi/10.1103/PhysRevB.100.134505},
  urldate = {2022-02-22}
}

@article{pengParityTimesymmetricWhisperinggallery2014,
  title = {Parity--Time-Symmetric Whispering-Gallery Microcavities},
  author = {Peng, Bo and {\"O}zdemir, {\c S}ahin Kaya and Lei, Fuchuan and Monifi, Faraz and Gianfreda, Mariagiovanna and Long, Gui Lu and Fan, Shanhui and Nori, Franco and Bender, Carl M. and Yang, Lan},
  year = {2014},
  month = may,
  journal = {Nature Physics},
  volume = {10},
  number = {5},
  pages = {394--398},
  publisher = {Nature Publishing Group},
  issn = {1745-2481},
  doi = {10.1038/nphys2927},
  url = {https://www.nature.com/articles/nphys2927},
  urldate = {2022-02-22},
  copyright = {2014 Nature Publishing Group},
  langid = {english}
}

@article{PhysRev.105.390,
  title = {Magnetostatic Modes in Ferromagnetic Resonance},
  author = {Walker, L. R.},
  year = {1957},
  month = jan,
  journal = {Phys. Rev.},
  volume = {105},
  number = {2},
  pages = {390--399},
  publisher = {American Physical Society},
  doi = {10.1103/PhysRev.105.390},
  url = {https://link.aps.org/doi/10.1103/PhysRev.105.390},
  numpages = {0}
}

@article{PhysRev.114.739,
  title = {Identification of the Magnetostatic Modes of Ferrimagnetic Resonant Spheres},
  author = {Fletcher, P. and Solt, I. H. and Bell, R.},
  year = {1959},
  month = may,
  journal = {Phys. Rev.},
  volume = {114},
  number = {3},
  pages = {739--745},
  publisher = {American Physical Society},
  doi = {10.1103/PhysRev.114.739},
  url = {https://link.aps.org/doi/10.1103/PhysRev.114.739},
  numpages = {0}
}

@article{PhysRevApplied.2.054002,
  title = {High-Cooperativity Cavity {{QED}} with Magnons at Microwave Frequencies},
  author = {Goryachev, Maxim and Farr, Warrick G. and Creedon, Daniel L. and Fan, Yaohui and Kostylev, Mikhail and Tobar, Michael E.},
  year = {2014},
  month = nov,
  journal = {Phys. Rev. Appl.},
  volume = {2},
  number = {5},
  pages = {054002},
  publisher = {American Physical Society},
  doi = {10.1103/PhysRevApplied.2.054002},
  url = {https://link.aps.org/doi/10.1103/PhysRevApplied.2.054002},
  numpages = {11}
}

@article{qianNonHermitianControlAbsorption2023,
  title = {Non-{{Hermitian}} Control between Absorption and Transparency in Perfect Zero-Reflection Magnonics},
  author = {Qian, Jie and Meng, C. H. and Rao, J. W. and Rao, Z. J. and An, Zhenghua and Gui, Yongsheng and Hu, C.-M.},
  year = {2023},
  month = jun,
  journal = {Nature Communications},
  volume = {14},
  number = {1},
  pages = {3437},
  publisher = {Nature Publishing Group},
  issn = {2041-1723},
  doi = {10.1038/s41467-023-39102-3},
  url = {https://www.nature.com/articles/s41467-023-39102-3},
  urldate = {2023-11-29},
  copyright = {2023 The Author(s)},
  langid = {english}
}

@article{qiFloquetGenerationMagnonic2023,
  title = {Floquet Generation of a Magnonic {{NOON}} State},
  author = {Qi, Shi-fan and Jing, Jun},
  year = {2023},
  month = jan,
  journal = {Physical Review A},
  volume = {107},
  number = {1},
  pages = {013702},
  publisher = {American Physical Society},
  doi = {10.1103/PhysRevA.107.013702},
  url = {https://link.aps.org/doi/10.1103/PhysRevA.107.013702},
  urldate = {2024-05-17}
}

@article{schumerTopologicalModesLaser2022,
  title = {Topological Modes in a Laser Cavity through Exceptional State Transfer},
  author = {Schumer, A. and Liu, Y. G. N. and Leshin, J. and Ding, L. and Alahmadi, Y. and Hassan, A. U. and Nasari, H. and Rotter, S. and Christodoulides, D. N. and LiKamWa, P. and Khajavikhan, M.},
  year = {2022},
  month = feb,
  journal = {Science},
  volume = {375},
  number = {6583},
  pages = {884--888},
  publisher = {American Association for the Advancement of Science},
  doi = {10.1126/science.abl6571},
  url = {https://www.science.org/doi/full/10.1126/science.abl6571},
  urldate = {2023-02-09}
}

@article{stehmannObservationExceptionalPoints2004,
  title = {Observation of Exceptional Points in Electronic Circuits},
  author = {Stehmann, T. and Heiss, W. D. and Scholtz, F. G.},
  year = {2004},
  month = jul,
  journal = {Journal of Physics A: Mathematical and General},
  volume = {37},
  number = {31},
  pages = {7813--7819},
  publisher = {IOP Publishing},
  issn = {0305-4470},
  doi = {10.1088/0305-4470/37/31/012},
  url = {https://doi.org/10.1088/0305-4470/37/31/012},
  urldate = {2021-03-29},
  langid = {english}
}

@article{tabuchiCoherentCouplingFerromagnetic2015,
  title = {Coherent Coupling between a Ferromagnetic Magnon and a Superconducting Qubit},
  author = {Tabuchi, Yutaka and Ishino, Seiichiro and Noguchi, Atsushi and Ishikawa, Toyofumi and Yamazaki, Rekishu and Usami, Koji and Nakamura, Yasunobu},
  year = {2015},
  month = jul,
  journal = {Science},
  volume = {349},
  number = {6246},
  pages = {405--408},
  publisher = {American Association for the Advancement of Science},
  doi = {10.1126/science.aaa3693},
  url = {https://www.science.org/doi/full/10.1126/science.aaa3693},
  urldate = {2023-01-25}
}

@article{tabuchiHybridizingFerromagneticMagnons2014,
  title = {Hybridizing {{Ferromagnetic Magnons}} and {{Microwave Photons}} in the {{Quantum Limit}}},
  author = {Tabuchi, Yutaka and Ishino, Seiichiro and Ishikawa, Toyofumi and Yamazaki, Rekishu and Usami, Koji and Nakamura, Yasunobu},
  year = {2014},
  month = aug,
  journal = {Physical Review Letters},
  volume = {113},
  number = {8},
  pages = {083603},
  publisher = {American Physical Society},
  doi = {10.1103/PhysRevLett.113.083603},
  url = {https://link.aps.org/doi/10.1103/PhysRevLett.113.083603},
  urldate = {2023-01-25}
}

@article{uzdinObservabilityAsymmetryAdiabatic2011,
  title = {On the Observability and Asymmetry of Adiabatic State Flips Generated by Exceptional Points},
  author = {Uzdin, Raam and Mailybaev, Alexei and Moiseyev, Nimrod},
  year = {2011},
  month = oct,
  journal = {Journal of Physics A: Mathematical and Theoretical},
  volume = {44},
  number = {43},
  pages = {435302},
  publisher = {IOP Publishing},
  issn = {1751-8121},
  doi = {10.1088/1751-8113/44/43/435302},
  url = {https://dx.doi.org/10.1088/1751-8113/44/43/435302},
  urldate = {2024-05-18},
  langid = {english}
}

@article{vitanovStimulatedRamanAdiabatic2017,
  title = {Stimulated {{Raman}} Adiabatic Passage in Physics, Chemistry, and Beyond},
  author = {Vitanov, Nikolay V. and Rangelov, Andon A. and Shore, Bruce W. and Bergmann, Klaas},
  year = {2017},
  month = mar,
  journal = {Reviews of Modern Physics},
  volume = {89},
  number = {1},
  pages = {015006},
  publisher = {American Physical Society},
  doi = {10.1103/RevModPhys.89.015006},
  url = {https://link.aps.org/doi/10.1103/RevModPhys.89.015006},
  urldate = {2024-05-17}
}

@article{wangEnhancementMagnonicFrequency2024,
  title = {Enhancement of Magnonic Frequency Combs by Exceptional Points},
  author = {Wang, Congyi and Rao, Jinwei and Chen, Zhijian and Zhao, Kaixin and Sun, Liaoxin and Yao, Bimu and Yu, Tao and Wang, Yi-Pu and Lu, Wei},
  year = {2024},
  month = apr,
  journal = {Nature Physics},
  pages = {1--6},
  publisher = {Nature Publishing Group},
  issn = {1745-2481},
  doi = {10.1038/s41567-024-02478-0},
  url = {https://www.nature.com/articles/s41567-024-02478-0},
  urldate = {2024-05-11},
  copyright = {2024 The Author(s), under exclusive licence to Springer Nature Limited},
  langid = {english}
}

@article{wangNonreciprocityUnidirectionalInvisibility2019,
  title = {Nonreciprocity and {{Unidirectional Invisibility}} in {{Cavity Magnonics}}},
  author = {Wang, Yi-Pu and Rao, J. W. and Yang, Y. and Xu, Peng-Chao and Gui, Y. S. and Yao, B. M. and You, J. Q. and Hu, C.-M.},
  year = {2019},
  month = sep,
  journal = {Physical Review Letters},
  volume = {123},
  number = {12},
  pages = {127202},
  publisher = {American Physical Society},
  doi = {10.1103/PhysRevLett.123.127202},
  url = {https://link.aps.org/doi/10.1103/PhysRevLett.123.127202},
  urldate = {2023-12-06}
}

@article{xuFloquetCavityElectromagnonics2020,
  title = {Floquet {{Cavity Electromagnonics}}},
  author = {Xu, Jing and Zhong, Changchun and Han, Xu and Jin, Dafei and Jiang, Liang and Zhang, Xufeng},
  year = {2020},
  month = dec,
  journal = {Physical Review Letters},
  volume = {125},
  number = {23},
  pages = {237201},
  publisher = {American Physical Society},
  doi = {10.1103/PhysRevLett.125.237201},
  url = {https://link.aps.org/doi/10.1103/PhysRevLett.125.237201},
  urldate = {2023-12-03}
}

@article{xuTopologicalEnergyTransfer2016,
  title = {Topological Energy Transfer in an Optomechanical System with Exceptional Points},
  author = {Xu, H. and Mason, D. and Jiang, Luyao and Harris, J. G. E.},
  year = {2016},
  month = sep,
  journal = {Nature},
  volume = {537},
  number = {7618},
  pages = {80--83},
  publisher = {Nature Publishing Group},
  issn = {1476-4687},
  doi = {10.1038/nature18604},
  url = {https://www.nature.com/articles/nature18604},
  urldate = {2024-01-18},
  copyright = {2016 Macmillan Publishers Limited, part of Springer Nature. All rights reserved.},
  langid = {english}
}

@article{yangTheoryFloquetdrivenDissipative2023,
  title = {Theory of {{Floquet-driven}} Dissipative Cavity Magnonics},
  author = {Yang, Ying and Xiao, Yang and Hu, C.-M.},
  year = {2023},
  month = feb,
  journal = {Physical Review B},
  volume = {107},
  number = {5},
  pages = {054413},
  publisher = {American Physical Society},
  doi = {10.1103/PhysRevB.107.054413},
  url = {https://link.aps.org/doi/10.1103/PhysRevB.107.054413},
  urldate = {2024-05-09}
}

@article{yoonTimeasymmetricLoopExceptional2018,
  title = {Time-Asymmetric Loop around an Exceptional Point over the Full Optical Communications Band},
  author = {Yoon, Jae Woong and Choi, Youngsun and Hahn, Choloong and Kim, Gunpyo and Song, Seok Ho and Yang, Ki-Yeon and Lee, Jeong Yub and Kim, Yongsung and Lee, Chang Seung and Shin, Jai Kwang and Lee, Hong-Seok and Berini, Pierre},
  year = {2018},
  month = oct,
  journal = {Nature},
  volume = {562},
  number = {7725},
  pages = {86--90},
  publisher = {Nature Publishing Group},
  issn = {1476-4687},
  doi = {10.1038/s41586-018-0523-2},
  url = {https://www.nature.com/articles/s41586-018-0523-2},
  urldate = {2024-01-19},
  copyright = {2018 Springer Nature Limited},
  langid = {english}
}

@article{zarerameshtiCavityMagnonics2022,
  title = {Cavity Magnonics},
  author = {Zare Rameshti, Babak and Viola Kusminskiy, Silvia and Haigh, James A. and Usami, Koji and {Lachance-Quirion}, Dany and Nakamura, Yasunobu and Hu, Can-Ming and Tang, Hong X. and Bauer, Gerrit E. W. and Blanter, Yaroslav M.},
  year = {2022},
  month = sep,
  journal = {Physics Reports},
  series = {Cavity {{Magnonics}}},
  volume = {979},
  pages = {1--61},
  issn = {0370-1573},
  doi = {10.1016/j.physrep.2022.06.001},
  url = {https://www.sciencedirect.com/science/article/pii/S0370157322002460},
  urldate = {2023-08-10},
  langid = {english}
}

@article{zhangBroadbandNonreciprocityEnabled2020,
  title = {Broadband {{Nonreciprocity Enabled}} by {{Strong Coupling}} of {{Magnons}} and {{Microwave Photons}}},
  author = {Zhang, Xufeng and Galda, Alexey and Han, Xu and Jin, Dafei and Vinokur, V. M.},
  year = {2020},
  month = apr,
  journal = {Physical Review Applied},
  volume = {13},
  number = {4},
  pages = {044039},
  publisher = {American Physical Society},
  doi = {10.1103/PhysRevApplied.13.044039},
  url = {https://link.aps.org/doi/10.1103/PhysRevApplied.13.044039},
  urldate = {2023-12-06}
}

@article{zhangExperimentalObservationExceptional2019,
  title = {Experimental {{Observation}} of an {{Exceptional Surface}} in {{Synthetic Dimensions}} with {{Magnon Polaritons}}},
  author = {Zhang, Xufeng and Ding, Kun and Zhou, Xianjing and Xu, Jing and Jin, Dafei},
  year = {2019},
  month = dec,
  journal = {Physical Review Letters},
  volume = {123},
  number = {23},
  pages = {237202},
  publisher = {American Physical Society},
  doi = {10.1103/PhysRevLett.123.237202},
  url = {https://link.aps.org/doi/10.1103/PhysRevLett.123.237202},
  urldate = {2023-12-06}
}

@article{zhangNonHermitianShortcutAdiabaticity2022,
  title = {Non-{{Hermitian}} Shortcut to Adiabaticity in {{Floquet}} Cavity Electromagnonics},
  author = {Zhang, Feng-Yang and Wu, Qi-Cheng and Yang, Chui-Ping},
  year = {2022},
  month = jul,
  journal = {Physical Review A},
  volume = {106},
  number = {1},
  pages = {012609},
  publisher = {American Physical Society},
  doi = {10.1103/PhysRevA.106.012609},
  url = {https://link.aps.org/doi/10.1103/PhysRevA.106.012609},
  urldate = {2024-05-17}
}

@article{zhangObservationExceptionalPoint2017,
  title = {Observation of the Exceptional Point in Cavity Magnon-Polaritons},
  author = {Zhang, Dengke and Luo, Xiao-Qing and Wang, Yi-Pu and Li, Tie-Fu and You, J. Q.},
  year = {2017},
  month = nov,
  journal = {Nature Communications},
  volume = {8},
  number = {1},
  pages = {1368},
  publisher = {Nature Publishing Group},
  issn = {2041-1723},
  doi = {10.1038/s41467-017-01634-w},
  url = {https://www.nature.com/articles/s41467-017-01634-w},
  urldate = {2021-03-29},
  copyright = {2017 The Author(s)},
  langid = {english}
}

@article{zhangPhononLaserOperating2018,
  title = {A Phonon Laser Operating at an Exceptional Point},
  author = {Zhang, Jing and Peng, Bo and {\"O}zdemir, {\c S}ahin Kaya and Pichler, Kevin and Krimer, Dmitry O. and Zhao, Guangming and Nori, Franco and Liu, Yu-xi and Rotter, Stefan and Yang, Lan},
  year = {2018},
  month = aug,
  journal = {Nature Photonics},
  volume = {12},
  number = {8},
  pages = {479--484},
  publisher = {Nature Publishing Group},
  issn = {1749-4893},
  doi = {10.1038/s41566-018-0213-5},
  url = {https://www.nature.com/articles/s41566-018-0213-5},
  urldate = {2024-01-29},
  copyright = {2018 The Author(s)},
  langid = {english}
}

@article{zhangStronglyCoupledMagnons2014,
  title = {Strongly {{Coupled Magnons}} and {{Cavity Microwave Photons}}},
  author = {Zhang, Xufeng and Zou, Chang-Ling and Jiang, Liang and Tang, Hong X.},
  year = {2014},
  month = oct,
  journal = {Physical Review Letters},
  volume = {113},
  number = {15},
  pages = {156401},
  publisher = {American Physical Society},
  doi = {10.1103/PhysRevLett.113.156401},
  url = {https://link.aps.org/doi/10.1103/PhysRevLett.113.156401},
  urldate = {2023-01-25}
}

@article{zhuFloquetengineeringMagnonicNOON2023,
  title = {Floquet-Engineering Magnonic {{NOON}} States with Performance Improved by Soft Quantum Control},
  author = {Zhu, Xinying and Xia, Ran and Xu, Liuyang},
  year = {2023},
  month = dec,
  journal = {Quantum Information Processing},
  volume = {22},
  number = {12},
  pages = {454},
  issn = {1573-1332},
  doi = {10.1007/s11128-023-04200-0},
  url = {https://doi.org/10.1007/s11128-023-04200-0},
  urldate = {2024-05-17},
  langid = {english}
}

@article{znojilPassageExceptionalPoint2020,
  title = {Passage through Exceptional Point: Case Study},
  shorttitle = {Passage through Exceptional Point},
  author = {Znojil, Miloslav},
  year = {2020},
  month = apr,
  journal = {Proceedings of the Royal Society A: Mathematical, Physical and Engineering Sciences},
  volume = {476},
  number = {2236},
  pages = {20190831},
  publisher = {Royal Society},
  doi = {10.1098/rspa.2019.0831},
  url = {https://royalsocietypublishing.org/doi/10.1098/rspa.2019.0831},
  urldate = {2024-07-05}
}

\end{document}